\documentstyle[floats,epsfig,twocolumn,prd,aps]{revtex}

\begin{document}\draft
%\setlength{\unitlength}{1mm}

%%%%%%%%%%%%%%%%%%%%%%%%%%%%%%%%%%%%%%%%%%%%%%%%%%%%%%%%%%%%%%%%%%%%%%%%
\newcommand{\be}{\begin{equation}}
\newcommand{\ee}{\end{equation}}
\newcommand{\ba}{\begin{eqnarray}}
\newcommand{\ea}{\end{eqnarray}}
\newcommand{\ban}{\begin{eqnarray*}}
\newcommand{\ean}{\end{eqnarray*}}

\newcommand{\e}{{\mathrm e}}
\newcommand{\ra}{\rangle}
\newcommand{\la}{\langle}

\newcommand{\n}[1]{\label{#1}}
\newcommand{\eq}[1]{Eq.(\ref{#1})}
\newcommand{\ind}[1]{\mbox{\bf\tiny{#1}}}
\renewcommand\theequation{\thesection.\arabic{equation}}

\newcommand{\nn}{\nonumber \\ \nonumber \\}
\newcommand{\nl}{\\  \nonumber \\}
\newcommand{\pr}{\partial}
\renewcommand{\vec}[1]{\mbox{\boldmath$#1$}}
%%%%%%%%%%%%%%%%%%%%%%%%%%%%%%%%%%%%%%%%%%%%%%%%%%%%%%%%%%%%%%%%%%%%%%%%

\wideabs{%%% Remove before submission!
\title{Stationary strings near a higher-dimensional rotating black hole}
\author{
Valeri P. Frolov$^*$ and 
Kory A. Stevens$^\dagger$
}
\address{
  \medskip
  Theoretical Physics Institute, Department of Physics,\\
  University of Alberta,  Edmonton,\\ Canada T6G 2J1\\
}
\address{
  \medskip  {$^*$\rm E-mail: \texttt{frolov@phys.ualberta.ca}}
  }
\address{
  \medskip  {$^\dagger$\rm E-mail: \texttt{kstevens@phys.ualberta.ca}}
  }

\maketitle

\begin{abstract}  
We study stationary string configurations in a
space-time of a higher-dimensional rotating black hole. We
demonstrate that the Nambu-Goto equations for a stationary string in
the 5D Myers-Perry metric allow a separation  of variables. We
present these equations  in the first-order form and study their
properties. We prove that the only stationary string configuration
which crosses the infinite red-shift surface and remains regular
there is a principal Killing string. A worldsheet of such a string is
generated by a principal null geodesic and a timelike at infinity
Killing vector field. We obtain principal Killing string solutions in
the Myers-Perry metrics with an arbitrary number of dimensions. It is
shown that due to the interaction of a  string with a rotating black
hole there is an angular momentum transfer from the black hole to the
string. We calculate the rate of this transfer in a spacetime with an
arbitrary number of dimensions.  This effect slows down the rotation
of the black hole. We discuss possible  final stationary
configurations of a rotating black hole interacting with a string.
\end{abstract}

\pacs{PACS numbers: 04.50.+h, 98.80.Cq \hfill 
 Alberta-Thy-08-04}
}
\narrowtext

\section{Introduction}
\setcounter{equation}0

It is well known that the Kerr metric possesses a number what was
called by Chandrasekhar \cite{Chandra} `miraculous' properties such
as the separability of the geodesic Hamilton-Jacobi equation
\cite{Cart:67}, separability of a scalar field equation
\cite{Cart:68}, separability and decoupling of the massless
non-zero-spin field equations \cite{Teuk:68} and separability of the
equilibrium equation for a stationary cosmic string
\cite{FrSkZeHe:89,CaFr:89}. These properties are closely connected
with the existence of the second order Killing tensor discovered in
the Kerr metric by Carter \cite{Cart:67}. 

A lot of attention has been focused recently on the study the higher
dimensional generalizations of the Kerr metric and their properties.
This interest is partly connected with brane world scenarios where
our physical world is represented by a 4-dimensional brane embedded
in the higher dimensional bulk spacetime. A higher dimensional
generalization of the metric of a rotating black hole was obtained by
Myers and Perry in 1986 \cite{MyPe:86}. It is remarkable that the
five dimensional Myers-Perry (MP) metric possesses `miraculous'
properties similar to the Kerr metric. Namely it allows a separation
of variables in the geodesic Hamilton-Jacobi equation \cite{FrSt:03a}
and the separability of the massless scalar field equation
\cite{FrSt:03b}. These properties are also connected with the
existence of the second order Killing tensor \cite{FrSt:03a}. 

In this paper we study stationary equilibrium configuration of a
cosmic string in the spacetime of a higher-dimensional rotating black
hole. We shall demonstrate that in the 5D spacetime the equation of
motion for such a string allows separation of variables. We shall
also show that in the five dimensional case there is an analogue of
the 4-dimensional `string uniqueness theorem' \cite{FrHeLa:96},
namely, the only stationary string configuration which crosses the
infinite red-shift surface and remains regular there is a principal
Killing string. A worldsheet of such a string is generated by 
principal null geodesics and the timelike at infinity Killing vector
field. We obtain the principal Killing string solutions in the
Myers-Perry metric with an arbitrary number of dimensions. And
finally, we  consider the interaction of a stationary string with a
higher-dimensional rotating black hole and demonstrate that in a
general case there exists an angular momentum  transfer from the
black hole to the string. As a result of this friction effect, the
black hole rotation is slowed down until the final stationary
configuration is reached. We discuss possible final stationary states
of such systems.

\section{A stationary string in a stationary spacetime}
\setcounter{equation}0

A general stationary metric in a spacetime $M$ with $N$ spatial
dimensions can be written in the form ($\mu,\nu=0\ldots N$)
\be\n{2.1}
ds^2=g_{\mu\nu}dx^{\mu}\, dx^{\nu}=
-F\, (dt+A_i\, dx^i)^2+  H_{ij}\, dx^i\, dx^j\, ,
\ee
where $F$, $A_i$, and $h_{ij}$ are functions of spatial coordinates
$x^i$ ($i,j=1\ldots N$). Denote by
$\xi^{\mu}\partial_{\mu}=\partial_t$ the Killing vector. Then
\be\n{2.2}
\xi^2=-F\, ,\hspace{0.5cm}\xi_{i}=-F\, A_i\, .
\ee
We assume that the spacetime is a $N$-dimensional foliation of the
Killing trajectories and denote ${\cal M}=M/G_1$ the factor space.
Elements of ${\cal M}$ are orbits of 1-dimensional group $G_1$
generated by $\xi$. A tensor
\be\n{2.3}
H_{\mu\nu}=g_{\mu\nu} +{\xi_{\mu}\, \xi_{\nu}\over F}\, 
\ee
is the projector onto the factor space ${\cal M}$.  It also possesses
the property
\be\n{2.4}
{\cal L}_{\xi}\, H =0\, ,
\ee
where ${\cal L}_{\xi}$ is the Lie derivative in the direction $\xi$. 

Denote by $H^{ij}$ an $N\times N$ matrix defined as
\be\n{2.5}
H_{ij}\, H^{jk}=\delta_i^k\, ,
\ee
then the contravariant components of the metric $g$ are
\be\n{2.6}
g^{00}=-F^{-1}+F^{-2}\, H^{ij}\, \xi_i\, \xi_j\, ,
\ee
\be\n{2.7}
g^{0i}=F^{-1}\, H^{ij}\, \xi_j\, ,\hspace{0.5cm}
g^{ij}=H^{ij}\, .
\ee

Any stationary tensor $T$, ${\cal L}_{\xi}T=0$, can be projected onto
${\cal M}$ by using the projector operator $H$. 
Geroch \cite{Geroch} demonstrated that the derivative of a stationary
tensor defined as
\be\n{2.8}
{T^{\ldots \mu}_{\nu\ldots}}_{:\lambda}=H^{\mu}_{\alpha}\cdots
H^{\beta}_{\nu}\, H^{\gamma}_{\lambda}\, {T^{\ldots
\alpha}_{\beta\ldots}}_{;\gamma}\, 
\ee
obeys all axioms of a covariant derivative connected with the metric
$H$.

Suppose that in addition to the Killing vector $\xi$ the spacetime $M$
has also either Killing vector $\eta^{\mu}$ and/or Killing tensor
$K^{\mu\nu}$, 
\be\n{2.9}
\eta_{(\mu;\nu)}=0\, ,\hspace{0.5cm}
K_{(\mu\nu;\lambda)}=0\, ,
\ee
which obey the conditions
\be\n{2.10}
{\cal L}_{\xi}\, \eta=0\, ,\hspace{0.5cm}
{\cal L}_{\xi}\, K=0\, .
\ee
Their projection onto ${\cal M}$ have components $\eta^i$ and $K^{ij}$
respectively.  It is easy to show (see \cite{CaFr:89}) that these
projections obey the equations
\be\n{2.11}
\eta_{(i:j)}=0\, ,\hspace{0.5cm}
K_{(ij:k)}=0\, ,
\ee
and hence they are a Killing vector and a Killing tensor for the
metric $H_{ij}$, respectively.

We call a 2-dimensional surface $\Sigma$ {\em stationary} if it is
tangent to $\xi$. A stationary surface is formed by a one dimensional
family  of the Killing trajectories for the field $\xi$. We choose
coordinates $\zeta^A$ ($A=0,1$) on $\Sigma$ so that $\zeta^0=t$ and
denote $\zeta^1=\sigma$. The embedding of $\Sigma$ into $M$ is
determined by its projection $x^{i}=x^{i}(\sigma)$ in ${\cal M}$. The
induced metric $G_{AB}$ on $\Sigma$ is
\be\n{2.12}
d\gamma^2=G_{AB}\, d\zeta^A\, d\zeta^B=-F\, (dt+{\cal A}\, d\sigma)^2+{\cal H}\, d\sigma^2\, ,
\ee
where
\be\n{2.13}
{\cal A}=A_{i}{dx^i\over d\sigma}\, ,\hspace{0.5cm}
{\cal H}=H_{ij}\,{dx^i\over d\sigma}\,{dx^j\over d\sigma}\, .
\ee 

A test string worldsheet equation of motion can be obtained as an
extremum of the Nambu-Goto action
\be\n{2.14}
I[x^{\mu}(\zeta^A)]=-\mu^* \, \int \sqrt{ -\det (G_{AB})}\, ,
\ee
where $\mu^*$ is the string tension and $G_{AB}$ is the induced metric.
For a stationary string this action reduces to
\be\n{2.15}
I=-\Delta t\, E\, ,
\ee
where
\be\n{2.16}
E=\mu^*\,\int d\sigma\, \sqrt{h_{ij}\,{dx^i\over d\sigma}\,{dx^j\over
d\sigma}}\, ,\hspace{0.5cm}
h_{ij}=F\, H_{ij}\, .
\ee
In other words, a line $x^{i}$ representing the stationary string in
the projection space ${\cal M}$ is a geodesic in the metric $h_{ij}$.
One can also interpret this result as follows. $dl=\sqrt{H_{ij}dx^i\,
dx^j}$ is the element of proper length of the string. The proper mass
of this element is $dm=\mu^* dl$, while its energy $dE$ differs from
$dm$ by a red-shift factor $\sqrt{F}$. The expression (\ref{2.16}) is
obtained by summing $dE$ over all the elements of the string.

\section{Stationary strings in 5-D MP geometry}
\setcounter{equation}0

We consider a stationary string  in a spacetime of a
5-dimensional rotating black hole. The corresponding MP metric 
\cite{MyPe:86} in the
Boyer-Lindquist coordinates is:
\be \n{3.1}  
ds^2=- dt^2  + {r_0^2\over \rho^2} \left[dt+a\, \sin^2\theta\, d\phi +b\,
\cos^2\theta\, d\psi  \right]^2
\ee
\[
+ (x+b^2)\,
\cos^2\theta\, d\psi^2 
+(x+a^2)\, \sin^2\theta\, d\phi^2+{\rho^2\over 4\Delta}\, dx^2+\rho^2d\theta^2\, .
\]
Here,
\be\n{3.2} 
\rho^2=x+P^2\,
,\hspace{0.3cm}P=\sqrt{a^2\,\cos^2\theta+b^2\,\sin^2\theta}\, ,
\ee
\be\n{3.3} 
\Delta=(x+a^2)(x+b^2)-r_0^2\, x\, .
\ee
The gravitational radius $r_0$ and the rotation parameters, $a$ and
$b$,  are connected with the mass $M$ of the
black hole and its angular momenta as follows
\be
M={3\pi\over 8}r_0^2\, ,
\ee
\be
J_{a}=-{2\over 3} Ma\, ,\hspace{0.2cm}J_{b}=-{2\over 3} Mb\, .
\ee
(Note that the sign of the rotation parameters in the MP metric is
opposite to the one adopted in the Kerr metric.) Angles $\phi$ and
$\psi$ take values from the interval $\left[0,2\pi \right]$, while
angle $\theta$ takes values from $\left[0,\pi/2 \right]$. Note also
that instead of the `radius' $r$ we use the coordinate $x=r^2$. This
will allow us to simplify calculations and make many of the
expressions more compact.

The black hole horizon is located at $x=x_+$ where
\be\n{3.4} 
x_{\pm}={1\over 2}\left[B\pm
\sqrt{B^2-4a^2b^2}\right]\, ,
\ee
where $B=r_0^2-a^2-b^2$. The angular velocities $\Omega_a$ and
$\Omega_b$ and the surface gravity $\kappa$ are
\be \n{3.5} 
\Omega_a={a\over x_+ +a^2}\, , \hspace{0.2cm} \Omega_b={b\over
x_+ +b^2}\, ,\hspace{0.2cm} \kappa= \left. {\partial_x\Delta\, \over r_0^2\,
\sqrt{x} }\right|_{x=x_+} \, . 
\ee
The infinite red-shift surface, which is an external boundary of the
ergosphere, is determined by the equation $\xi^2=0$, or
$\rho^2=r_0^2$.

The metric (\ref{3.1}) is invariant under the following
transformation
\be \n{3.6} 
a \leftrightarrow b\, ,\hspace{0.5cm} \theta \leftrightarrow
\left( {\pi\over 2}-\theta \right)\, ,\hspace{0.5cm} \phi
\leftrightarrow \psi \, . 
\ee
It possesses 3 Killing vectors, $\partial_t$, $\partial_{\phi}$
and $\partial_{\psi}$. For $a=b$ the metric has 2 additional Killing
vectors  \cite{CGL,FrSt:03a}:
\be \n{3.7} 
\cos\, \partial_{\bar{\theta}}-\cot\bar{\theta}\, \sin\bar{\phi}\,
\partial_{\bar{\phi}}+{\sin \bar{\phi}\over \sin\bar{\theta}}\,
\partial_{\bar{\psi}}\, ,
\ee
and
\be \n{3.8} 
-\sin\bar{\phi}\, \partial_{\bar{\theta}}-\cot\bar{\theta}\,
\cos\bar{\phi}\,
\partial_{\bar{\phi}}+{\cos\bar{\phi}\over \sin\bar{\theta}}\,
\partial_{\bar{\psi}}\, ,
\ee
where $\bar{\phi}=\psi-\phi$, $\bar{\psi} =\psi+\phi $ and $\bar{\theta} 
=2 \theta$

Besides the Killing vectors the metric (\ref{3.1}) has also the Killing
tensor $K^{\mu\nu}$ \cite{FrSt:03a}
\[
K^{\mu\nu} =-P^2(g^{\mu\nu}+
\delta^\mu_t \delta^\nu_t)+\delta^\mu_\theta
   \delta^\nu_\theta
\]
\be  \n{3.9}     
+\frac{1}{\sin^2\theta}\delta^\mu_\phi
   \delta^\nu_\phi+\frac{1}{\cos^2\theta}
   \delta^\mu_\psi \delta^\nu_\psi\, .
\ee

By representing the metric (\ref{3.1}) in the form (\ref{2.1}) one
gets
\be \n{3.10} 
F=-\xi^2={C\over \rho^2}\, ,\hspace{0.5cm}
C=\rho^2-r_0^2\, ,
\ee
\[
H_{xx}={\rho^2\over{4\Delta}}\, ,\hspace{0.2cm}
H_{\theta\theta}=\rho^2\, ,\hspace{0.2cm}
H_{\phi\psi}={ab\sin^2\theta\cos^2\theta r_0^2 \over{C}}
\]
\be\n{3.11}
H_{\phi\phi}=\left(x+a^2+{r_0^2 a^2 \sin^2 
\theta\over{C}}\right)\sin^2\theta\, ,
\ee
\[
H_{\psi\psi}=\left(x+b^2+{r_0^2 b^2 \cos^2 
\theta\over{C}}\right)\cos^2\theta\, ,
\]
and obtains the following expression for $h^{ij}$
\[
h^{\phi\phi}=\frac{1}{C} \left[{1\over
\sin^2\theta}-{(a^2-b^2)(x+b^2)+b^2r_0^2\over \Delta}\right] \, ,
\]
\be \n{3.12} 
h^{\psi\psi}=\frac{1}{C} \left[{1\over
\cos^2\theta}+{(a^2-b^2)(x+a^2)-a^2r_0^2\over \Delta} \right]\, ,
\ee
\[
h^{\phi \psi}=-\frac{abr_0^2}{C \Delta}\, ,\hspace{0.2cm}
h^{xx}=4\frac{\Delta}{C}\, ,\hspace{0.2cm}
h^{\theta \theta}=\frac{1}{C}\, .
\]
We used GRTensor program for the calculations.

To study the geodesics in the metric (\ref{3.1}) it is convenient to
use the Hamilton-Jacobi method (see e.g. \cite{Cart:67,MTW}). The
corresponding Hamilton-Jacobi equation reads
\be\n{3.13} 
\frac{\partial S}{\partial\sigma} +
\frac{1}{2} h^{ij} \frac{\partial S}{\partial x^i}
\frac{\partial S}{\partial x^j}=0\, .
\ee
For the metric (\ref{3.12}) this equation allows separation of
variables
\be\n{3.14}
S= -\frac{1}{2}m^2 \sigma + \Phi\phi +
\Psi\psi + S_\theta + S_x 
\ee 
where $S_\theta$ and $S_x$ are functions of $\theta$ and $x$
respectively. Substituting this expression into
(\ref{3.13}), one obtains
\be\n{3.15}
\left({\partial S_\theta \over{\partial\theta}}\right)^2
-m^2 P^2+
{1\over{\sin^2\theta}}\Phi^2+{1\over{\cos^2\theta}}\Psi^2=K \, ,
\ee
and
\[
4\Delta\left({\partial S_x\over{\partial x}}\right)^2+m^2 (r_0^2-x)
-{r_0^2(x+a^2)(x+b^2)\over{\Delta}}{\mathcal E}^2
\]
\be\n{3.16}
-(a^2-b^2)\left({\Phi^2\over{x+a^2}}-{\Psi^2\over{x+b^2}}\right)=-K\, ,
\ee
where
\be\n{3.17}
{\mathcal E}={a\Phi\over{x+a^2}}+{b\Psi\over{x+b^2}}\, .
\ee
Here $K$ is a separation constant. These equations are similar to the
the separated equations for the motion of a particle in 5D MP metric
\cite{FrSt:03b}, with the following three differences. First, the
energy $E$ has been set to zero, second the mass squared $m^2$ has
been changed to $-m^2$ and third, there  is an extra term of $m^2
r_0^2$ in the $x$-equation. The parameter $m$ depends on the choice
of the length parameter $\sigma$. After derivation of the equations
for a stationary string we put $m=1$ and use a proper length in the
metric $h$ as a
parameter $\sigma$. The extra term  $m^2 r_0^2$ is connected with
presence of  red-shift factor $\sqrt{F}$ in (\ref{2.16}).

We can write equations (\ref{3.15}) and (\ref{3.16}) as
\be\n{3.18}
{\partial S_\theta \over{\partial\theta}}=
\sigma_\theta\sqrt{\Theta}\, ,\hspace{0,3cm}
{\partial S_x \over{\partial x}}=\sigma_x{\sqrt{X}\over{2\Delta}},
\ee
with $\Theta$ and $X$ given as
\be\n{3.19}
\Theta=K+m^2 P^2-{\Phi^2\over{\sin^2\theta}}
-{\Psi^2\over{\cos^2\theta}}\, ,
\ee
\[
X=\Delta\left[m^2(x-r_0^2)-K\right]
+r_0^2 (b\Phi +a\Psi)^2
\]
\be\n{3.20}
+(a^2-b^2)\left[\Phi^2 (x+b^2)-\Psi^2 (x+a^2)\right]\, .
\ee
The sign functions $\sigma_\theta =\pm$ and $\sigma_x =\pm$ in the
two equations are independent of each other. In each equation the
change of sign occurs when the expression on the right hand side
vanishes.

We can now write the Hamilton-Jacobi action as
\be\n{3.21}
S= -\frac{1}{2}m^2 \sigma + \Phi\phi + \Psi\psi + 
\sigma_\theta\int^\theta \sqrt{\Theta}d\theta
+\sigma_x \int^x {\sqrt{X}\over{2\Delta}}dx\, .
\ee
By setting the derivatives of $S$ with respect to $K$, $m^2$, $\Phi$,
$\Psi$ equal to zero, we get a solution for the Hamilton-Jacobi
equations in the following form
\be\n{3.22}
\int^\theta {d\theta\over{\sqrt\Theta}}=\int^x {dx\over{2\sqrt X}}\, ,
\ee
\be\n{3.23}
\sigma=\int^\theta
{P^2\over{\sqrt\Theta}}\, d\theta+\int^x
{x-r_0^2\over{2\sqrt X}}dx\, ,
\ee
\[
\phi=\int^\theta {\Phi\over{\sin^2\theta \sqrt\Theta}}d\theta
\]
\be\n{3.24}
-\int^x {1\over{\sqrt X}}\left[{ar_0^2 (x+b^2)\over{2\Delta}}{\mathcal E}
+{(a^2-b^2)\Phi\over{2(x+a^2)}}\right]dx\, ,
\ee
\[
\psi=\int^\theta {\Psi\over{\cos^2\theta \sqrt\Theta}}d\theta
\]
\be\n{3.25}
-\int^x {1\over{\sqrt X}}\left[{br_0^2 (x+a^2)\over{2\Delta}}{\mathcal E}
-{(a^2-b^2)\Psi\over{2(x+b^2)}}\right]dx\, .
\ee
By differentiating these equations with respect to $\sigma$, one
obtains the first order differential equations
\be\n{3.26}
C\, \dot{x}=\sigma_x 2\sqrt{X}\, ,
\ee
\be\n{3.27}
C\dot{\theta}=\sigma_\theta \sqrt\Theta\, ,
\ee
\be\n{3.28}
C\, \dot{\phi}={\Phi\over{\sin^2\theta}}
-{ar_0^2(x+b^2)\over{\Delta}}{\mathcal
E}-{(a^2-b^2)\Phi\over{x+a^2}}\, ,
\ee
\be\n{3.29}
C\, \dot{\psi}={\Psi\over{\cos^2\theta}}
-{br_0^2(x+a^2)\over{\Delta}}{\mathcal
E}+{(a^2-b^2)\Psi\over{x+b^2}}\, .
\ee
Here as earlier $C=\rho^2-r_0^2$ and a dot means a derivative with
respect to $\sigma$. At this point, $m$ is no longer needed. We put
$m=1$, so that $\sigma$ is the proper length in the metric $h$. 

The string equations contain only one scale parameter, $r_0$. By
simple rescaling one can always put $r_0=1$. To simplify relations in
the following three sections we shall use this choice. In order to
return back to the dimensional units one must make the following
changes
\[
x\to x/r_0^2\, ,\hspace{0.2cm}
a\to a/r_0\, ,\hspace{0.2cm}
b\to b/r_0\, ,
\]
\be\n{3.30}
\Phi\to \Phi r_0\, ,\hspace{0.2cm}
\Psi\to \Psi r_0\, ,\hspace{0.2cm}
K\to K r_0^2\, ,\hspace{0.2cm}
\sigma\to \sigma/r_0\, .
\ee

\section{Types of String Configurations}
\setcounter{equation}0

\subsection{Properties of the radial equation}

In a 5D spacetime if one starts with the variation problem
(\ref{2.16}), one obtains 4 second order geodesic equations which
require $4\times 2$ initial data. The integrals of motion $\Phi$,
$\Psi$, and $K$ and the normalization condition $h_{ij}\dot{x}^i
\dot{x}^j=1$ allow one to determine the initial data for $\dot{x}^i$.
Hence the integrals of motion and 4 initial conditions for $x^i$
uniquely specify a string configuration. Because of the symmetry of
the problem, two of these quantities $\phi(0)$ and $\psi(0)$ are
cyclic. Thus a stationary string in the 5D MP metric is completely
determined by $\Phi$, $\Psi$, and $K$ and $x(0)$ and $\theta(0)$. 

We discuss first properties of the radial equation (\ref{3.26}). For
given values of other parameters, $X$ given by (\ref{3.20})
considered as a function of $x$ is a third order polynomial. 
The
allowed configurations must only occur where $X\ge 0$, $X=0$ gives
radial turning points.
In the limit of large $x$, $X$ has a leading positive term $x^3$, so
configurations can extend to infinity. By integrating the equations
(\ref{3.26})--(\ref{3.29}) in the region $x\to\infty$ one obtains 
\be\n{4.1}
\sigma\sim \sqrt{x} \sim r\, ,\hspace{0.5cm}
\theta\sim \theta_0-{\Theta(\theta_0)\over r}\, ,
\ee
\be\n{4.2}
\phi\sim \phi_0-{\Phi\over r\sin^2\theta_0}\, ,\hspace{0.5cm}
\psi\sim \psi_0-{\Psi\over r\cos^2\theta_0}\, . 
\ee

$X$  as a function of $x$ either is a monotonically growing function,
or it has one local maximum and one local minimum at $x^-$ and $x^+$,
respectively. By solving the equation $dX/dx=0$ one finds
\be\n{4.3}
x^{\pm}={1\over 3}\left[ 2-a^2-b^2+K\pm \sqrt{{\cal B}}\right]\, .
\ee
Here
\[
{\cal
B}=K^2+(a^2+b^2+1)K-3(a^2-b^2)(\Phi^2-\Psi^2)
\]
\be\n{4.4}
+(a^2-b^2)^2+(1-a^2)(1-b^2)\, .
\ee
For negative value of ${\cal B}$ the function $X$ is monotonic. 
Let us denote 
\be\n{4.5}
K_{\pm}={1\over 2}\left[-(1+a^2+b^2)\pm \sqrt{{\cal C}}\right]\, ,
\ee
\be\n{4.6}
{\cal C}=3\left\{
4(a^2-b^2)(\Phi^2-\Psi^2)-[(a+b)^2-1][(a-b)^2-1]   
\right\} \, .
\ee
If ${\cal C}>0$ then ${\cal B}$ is positive at $K<K_-$ and $K>K_+$. If
${\cal C}<0$, ${\cal B}>$ for all values of $K$.  

The value of $X$ at the 5D black hole horizon (where $\Delta=0$) is 
\be\n{4.7} 
X(x_+)= x_+(\Phi\Omega_a+\Psi\Omega_b)^2\, . 
\ee

\subsection{Properties of the $\theta$ equation}

Let us examine the $\theta$ equation. The string can extend to the
subspace $\theta=0$ only if $\Phi=0$, and can reach the subspace
$\theta={\pi\over{2}}$ only if $\Psi=0$. For $\Psi=0$ the
configuration is in the $\theta={\pi\over{2}}$ plane only if
$K=\Phi^2- b^2$. Similarly, for $\Phi=0$ the configuration is in the
$\theta=0$ plane if $K=\Psi^2- a^2$.

Consider a special type of configuration when strings are aligned so
that $\theta$ remains constant $\theta=\theta_0$. This configuration
can occur when
\be\n{4.8}
\Theta(\theta_0)={d\Theta\over{d\theta}}(\theta_0)=0\, .
\ee
This is equivalent to
\be\n{4.9}
K+P^2-{\Phi^2\over{\sin^2\theta_0}}
-{\Psi^2\over{\cos^2\theta_0}}=0\, ,
\ee
\be\n{4.10}
{\Phi^2\over{\sin^4\theta_0}}-{\Psi^2\over{\cos^4\theta_0}}-(a^2-b^2)=0\,
,
\ee
Where for the second equation, we have excluded cases $\theta_0=0$
and $\theta_0=\pi /2$.

\subsection{A string within a brane}

Higher dimensional black holes are of special interest in the brane
world models where a physical world is represented by a
(3+1)-dimensional brane embedded in a bulk space with large or
infinite extra dimensions.  In these models the usual matter (bosons,
fermions and gauge fields) are localized on the brane. A cosmic
string formed from this matter must also be located on the brane. 
Let us assume that a spacetime has 1 spatial extra dimension and its
size is much larger than the gravitational radius of the black hole.
Then a stationary cosmic string interacting with such a 5D rotating
black hole attached to the brane is described by a special solutions
of the equations (\ref{3.26})-(\ref{3.29}). Let us consider this case
in more details.

As a result of the black-hole--brane interaction, in the presence of
a (3+1)-brane a stationary black hole can have only one parameter of
the rotation (this follows from the general analysis \cite{FFS}). We
put $b=0$ and choose $\psi=$const as the brane equation. The metric
on the brane can be obtained from the 5D MP metric
(\ref{3.1})-(\ref{3.3}) by putting $b=0$ and $\psi=$const there. In
order to preserve the latter condition one can also put $\Psi=0$ in
the string equations (\ref{3.26})-(\ref{3.29}). As a result one
obtains the following equations
\be\n{4.11}
q^2 \dot{x}=2\sigma_x \sqrt{{\cal X}}\, ,
\ee
\be\n{4.12}
q^2 \dot{\theta}=\sigma_{\theta}\sqrt{Q}\, ,
\ee
\be\n{4.13}
q^2\dot{\phi}=\Phi\left( {1\over \sin^2\theta}-{a^2\over
x-1+a^2}\right)\, ,
\ee
where $q^2=x+a^2\cos^2\theta -1$, and
\[
{\cal X}=x\left[ (x-1+a^2)(x-1-K)+a^2\Phi^2\right] \, ,
\]
\be\n{4.14}
Q=K+a^2\cos^2\theta-{\Phi^2\over \sin^2\theta}\, .
\ee

For a string lying in the equatorial plane $\theta=\pi/2$ these
equations allow further simplification. Since $Q=0$, $K=\Phi^2$, and
\be\n{4.15}
\dot{x}=2\sigma_x V\, ,\hspace{0.5cm}V=\sqrt{{x(x-1+a^2-\Phi^2)\over
x-1}}\, ,
\ee
\be\n{4.16}
\dot{\phi}={\Phi\over x-1+a^2}\, .
\ee
From these equations we have
\be\n{4.17}
{d\phi\over dx}={\sigma_x\over 2}{\Phi\over x-1+a^2}\sqrt{{x-1\over
x(x-1+a^2-\Phi^2)}}\, . 
\ee

Outside the infinite red-shift surface, where $x>1$, this equation may have
a singular point only if $|\Phi|>a$. This point $x=1-a^2+\Phi^2$ is a
turning point where the string has a minimal distance to the black
hole. For $|\Phi|=a$ this singular point disappears. For this value of
$\Phi$ the string crosses the ergosphere and enters the 5D horizon. 

In the case when $|\Phi|<a$, the string is not regular at the infinite
red-shift surface. To demonstrate this consider the induced metric on
the string worldsheet 
\[
d\gamma^2=\left[{x D^2-\Phi^2 a^2 \over{4xD^2
(D-\Phi^2)}}\right]dx^2-dt^2
\]
\be\n{4.18}
+{1\over{x}}\left[dt+{a\Phi\over{2D}}\sqrt{{x-1
\over{x(D-\Phi^2)}}}dx \right]^2\, ,
\ee
where we have defined $D=x-1+a^2$ for brevity. 
The 2D curvature has only one component which for the metric
(\ref{4.18}) is
\be\n{4.19}
R={2\over{x^2(x-1)^2}}[3(x-1)^2+(4x-3)(a^2-\Phi^2)]\, .
\ee
For $|\Phi|< a$ the curvature is infinite at the infinite redshift
surface, $x=1$. The curvature remains finite at  $x=1$  in a special
case $|\Phi|=a$ when it reduces to $R=6/x^2$.  To examine the nature
of the singularity for $|\Phi|< a$ let us consider the determinant of
the metric (\ref{4.18})
\be\n{4.20}
g=-{x-1 \over{4x(x+a^2-1-\Phi^2)}}.
\ee
We see that for values of $x>1$ the metric has a negative
determinant, signaling one positive and one negative eigenvalue,
corresponding to one spacelike and one timelike dimension. For
$|\Phi|=a$ it remains negative for $0<x<1$. 
However, for $|\Phi|<a$ the determinant is positive between 
$x=1$ and  $x=1-a^2+\Phi^2$, so that that induced metric has Euclidean
signature. The corresponding spacelike surface does not represent any
solution for a physical cosmic string.

In the next section we obtain a general solution for a stationary
string which enters ergosphere and horizon of 5D rotating black hole
and remains regular and timelike there. In the appendix we prove the
uniqueness of such a solution.

\section{Stationary strings attached to a rotating black hole}
\setcounter{equation}0

\subsection{Principal null rays}

We discuss now conditions when a stationary string can cross the
horizon and enter the black hole. In  a four dimensional case a
uniqueness theorem was proven \cite{FrHeLa:96}  according to which a
worldsheet of such a string must be generated by the timelike at
infinity Killing vector $\xi=\xi_{(t)}$ and a principal null vector
field $l$. We demonstrate that a similar result is valid in a 5D
case. 

The principal null vectors in 5D MP metric are defined as a solution
of the equation
\be\n{5.1}
{l_{\pm}}_{[\alpha}\, C_{\beta]\gamma\delta \epsilon}\,
{l_{\pm}}^{\gamma}\,l_{\pm}^{\delta}\, =0\, ,
\ee
where $C_{\beta\gamma\delta \epsilon }$ is the Weyl tensor. They are
of the form \cite{MyPe:86,FrSt:03b}
\[
l_{\pm}^{\mu}\partial_{\mu}= \pm 2\sqrt{x}\partial_x
\]
\be\n{5.2}
+{(x+a^2)(x+b^2)\over \Delta}\left[ \partial_t
-{a\over x+a^2}\partial_{\phi} -{b\over x+b^2}\partial_{\psi}\right]\, .
\ee
The integral lines of $l_{\pm}^{\mu}$ are geodesics.
By  analogy with similar congruences in the four-dimensional Kerr
geometry, we call the congruences generated by $l_{\pm}^{\mu}$ {\em
principal null congruences}.

One can define a convenient basis by accompanying the two null vectors
$l_+$ and $l_-$ by the vectors $m$, $\bar{m}$ and $k$ defined as
follows
\be\n{5.3} 
m^{\mu}\partial_{\mu}={1\over{\rho\sqrt{2}}}
( \partial_{\theta}+{i\sin\theta\cos\theta\over{P}}\not\partial )\, , 
\ee
\be\n{5.4}
\not\partial =(b^2-a^2)\partial_{t}
+{a\over{\sin^2\theta}}\partial_{\phi}
-{b\over{\cos^2\theta}}\partial_{\psi}\, ,
\ee
\be\n{5.5}
k^{\mu}\partial_{\mu}=
{1\over{\sqrt{x}P}}
\left( ab\partial_{t}-b\partial_{\phi}-a\partial_{\psi} \right)\, .
\ee
These vectors obey the following normalization conditions
\[
(m\cdot m)=(\bar{m}\cdot\bar{m})=0\, ,
\hspace{0.2cm}(m\cdot\bar{m})=1\,,\hspace{0.2cm}(l_{\pm}\cdot m)=0\, ,
\]
\be\n{5.6}
(k\cdot k)=1\,,\hspace{0.2cm}
(k\cdot m)=(k\cdot \bar{m})=(k\cdot l_{\pm})=0\, ,
\ee
\[
(l_+\cdot l_-)= -2x\rho^2/\Delta \, ,
\hspace{0.4cm}(l_{\pm}\cdot \xi) =-1\, .
\] 
Direct calculations (by using GRTensor) allow one to prove that for
this choice of the basis one has
\be\n{5.7}
\xi_{\mu\,;\nu}=-{\Delta F_{,x}\over{\rho^2\sqrt{x}}}l_{+[\mu}l_{-\nu]}
-{2i P(1-F)\over{\rho^2}}m_{[\mu}\bar{m}_{\nu]}\, .
\ee 
The relation (\ref{5.7}) shows that the principal null vectors
$l_{\pm}$ are eigenvectors of $\xi_{\mu\,;\nu}$
\be\n{5.8}
\xi_{\mu\,;\rho}l^{\rho}_{\pm}=\pm \beta l_{\pm \mu}\, ,
\hspace{0.2cm}
\beta=\sqrt{x}F_{,x}={1\over 2}F_{,r}\, .
\ee

\subsection{Principal Killing surfaces}

We use now vectors $\xi$ and $l$ to generate a stationary 2D surface.
We would like to have a surface which is regular at the future event
horizon $H^+$. For this reason we shall use the incoming principal
null vector $l_-$ which is linearly independent from $\xi$ at $H^+$
and denote it simply by $l$. The time symmetry implies that ${\cal
L}_{\xi}l=0$ where ${\cal L}_{\xi}$ is a Lie derivative with respect
to the vector field $\xi$. (For briefness we omit the index $t$).
This condition can be written as
\be\n{5.9}
[\xi,l]=0\, .
\ee
By the Frobenious theorem the relation (\ref{5.9}) implies that there
exists a 2D surface $\Sigma$ given by the equation
$x^{\mu}=x^{\mu}(\zeta)$, ($\zeta^A=(v,\lambda)$) such that
$x^\mu_{,0} = \xi^\mu$, $x^\mu_{,1} = -l^\mu$. We call $\Sigma$ a
principal Killing surface. The parameter $v$ coincides with the
Killing time, while $\lambda$ is an affine parameter along a
principal null geodesics.  In these coordinates the metric on 
$\Sigma$  takes the form
\be\n{5.10}
d\gamma^2=G_{AB}d\zeta^A d\zeta^B=-Fdv^2+2\alpha dv \,d\lambda\, ,
\ee
\be\n{5.11}
F=-\xi^2\, ,\hspace{0.5cm} \alpha=-(\xi\cdot l)\, .
\ee

Let us show that $\alpha$ is constant\footnote{In fact this result is
valid for any stationary 2D surface, see Appendix.} on $\Sigma$.
Really
\be\n{5.12}
{d\alpha\over
d\lambda}=-l^{\mu}(\xi^{\nu}l_{\nu})_{;\mu}=\xi_{\nu;\mu}l^{\nu}l^{\mu}
+\xi_{\nu}\, l^{\mu}l^{\nu}_{\, ;\mu}=0\, .
\ee
The first term in the right hand side vanishes since $\xi_{\mu;\nu}$
is antisymmetric, the second vanishes since the null congruence is
geodesic. Similarly
\be\n{5.13}
{d\alpha\over
dv}=-\xi^{\mu}(\xi^{\nu}l_{\nu})_{;\mu}=-\xi^{\mu}(l^{\nu}\xi_{\nu;\mu}
+\xi^{\nu} l_{\nu;\mu})=0\, . 
\ee
The latter equality follows from (\ref{5.9}). By rescaling the
affine parameter $\lambda$ one can put the constant $\alpha$ to be
equal to 1. We shall use this gauge.

We introduce three vectors, $n^\mu_R$ ($R=2,3,4$) 
orthogonal to the worldsheet $\Sigma$
\be\n{5.14} 
g_{\mu \nu}n^\mu_R n^\nu_S =\delta_{RS}\, ,
\hspace{0.3cm} g_{\mu \nu}x^\mu_{,A} n^\nu_R =0\,  
\ee 
so that they together with $\xi$ and $l$ form a complete set. For this
set the following relation is valid
\be\n{5.15} 
g^{\mu\nu}=G^{AB}x^\mu_{,A}
x^\nu_{,B}+\delta^{RS}n^\mu_R n^\nu_S\, . 
\ee 
The concrete form of these normal vectors $n_R$ is not important.

Consider the second fundamental form for the worldsheet $\Sigma$,
defined as 
\be\n{5.16}
\Omega_{RAB}=g_{\mu\nu}n^\mu_R x^\rho_{,A} (x^\nu_{,B}\,)_{;\rho}\, .
\ee 
For the worldsheet to be minimal, we will need to have a
vanishing trace of the second fundamental form 
\be\n{5.17}
\Omega_{R}\equiv G^{AB}\Omega_{RAB}=(n_{R}\cdot z)\, ,
\ee 
where 
\be \n{5.18}
z^{\nu}=G^{AB}x^{\rho}_{,A}\, (x^\nu_{,B}\,)_{;\rho}\, . 
\ee 
Simple calculations using (\ref{5.9}) give 
\be \n{5.19}
z^{\nu}=-2\xi^{\nu}_{\,\,;\rho} l^{\rho}+F l^{\rho} l^{\nu}_{\,\,;\rho}\, .  
\ee 
Since the integral lines of $l$ are
geodesics, the second term on the right hand side vanishes. 
Using (\ref{5.8}) we can write (\ref{5.19}) in the form
\be\n{5.20}
z^{\mu}=F_{,r} l^{\mu}\, .
\ee
Since $(n_R\cdot l)=0$ one has $\Omega_R=0$. This result implies that
the principal Killing surface $\Sigma$ is a minimal surface and hence
it is a stationary solution of the Nambu-Goto equations. This
solution describes a stationary  string which  crosses  the
ergosphere and enters the event horizon. It is shown in the appendix
that the principal Killing surface is the only stationary string
solution which crosses the infinite red-shift surface and remains a
regular time-like surface there.

\subsection{Explicit form of a solution for a principal Killing
string}

The expression (\ref{5.2}) for the principal null vectors implies that 
\be\n{5.21}
p={(x+a^2)(x+b^2)\over \Delta}\xi-l\, 
\ee
is a spacelike vector tangent to the string and one has
\be\n{5.22}
\dot{x}^i=p^i\, .
\ee
Thus the string equations (\ref{3.26})-(\ref{3.29}) for this case
take the form $\theta=\theta_0=$const,
\be\n{5.23}
\dot{x}=\mp 2\sqrt{x}\, ,
\hspace{0.3cm}
\dot{\phi}={a(x+b^2)\over \Delta}\, ,
\hspace{0.3cm}
\dot{\psi}={b(x+a^2)\over \Delta}\, .
\ee
Equation (\ref{5.23}) shows that $x=r^2=\sigma^2$.
By comparing these equations with (\ref{3.26})-(\ref{3.29}) one
obtains
\be\n{5.24}
\Phi=a\sin^2\theta_0\, ,\hspace{0.3cm}\Psi=b\cos^2\theta_0\, ,
\ee
\be\n{5.25}
K=(a^2-b^2)(\sin^2\theta_0-\cos^2\theta_0)\, .
\ee

The remaining two equations of motion 
\be\n{5.26}
\dot{\phi}={a(r^2+b^2)\over{(r^2+a^2)(r^2+b^2)-r^2}}\,,
\ee
\be\n{5.27}
\dot{\psi}={b(r^2+a^2)\over{(r^2+a^2)(r^2+b^2)-r^2}}\, ,
\ee
can be integrated and give
\[
\phi=\phi_0+{a\over{2(r_+^2-r_-^2)}}\left[{r_+^2+b^2 \over{r_+}}
\ln\left({r-r_+ \over{r+r_+}}\right)
\right.
\]
\be\n{5.28}
\left.
-{r_-^2+b^2 \over{r_-}}
\ln\left({r-r_- \over{r+r_-}}\right)\right] \, ,
\ee
\[
\psi=\psi_0+{b\over{2(r_+^2-r_-^2)}}\left[{r_+^2+a^2 \over{r_+}}
\ln\left({r-r_+ \over{r+r_+}}\right)
\right.
\]
\be\n{5.29}
\left.
-{r_-^2+a^2 \over{r_-}}
\ln\left({r-r_- \over{r+r_-}}\right)\right]\, ,
\ee
with $\phi_0$ and $\psi_0$ being initial data for the string,  and
$r_{\pm}$ being the horizon locations, defined in (\ref{3.4}). The
last two relations show that when $a\ne 0$ and $b\ne 0$, the
principal Killing string makes infinite number of turns in $\phi$ and
$\psi$ directions before it reaches the horizon $r_+$. This is a pure
coordinate effect. In the next section we show that in the ingoing
Eddington-Finkelstein coordinates which are regular at the horizon
there is no infinite winding.

\section{The principal Killing string as a 2D black hole}
\setcounter{equation}0

To study stationary strings in the  5D black hole exterior  we used
the MP metric in the Boyer-Lindquist coordinates (\ref{3.1}). The
principal Killing strings cross infinite red-shift surface and enter
the 5D black hole horizon. For studying global properties of these
solutions instead of Boyer-Lindquist coordinates it is more
convenient to use the  ingoing Eddington-Finkelstein coordinates
which are regular at the future event horizon. The corresponding
coordinate transformation is
\be\n{6.1}
dv=dt+  {(x+a^2)(x+b^2)\over{2\Delta\sqrt{x}}}dx\, ,
\ee
\be\n{6.2}
d\tilde{\phi}=d\phi -{(x+b^2)a\over{2\Delta\sqrt{x}}}dx\, ,
\ee
\be\n{6.3}
d\tilde{\psi}=d\psi -{(x+a^2)b\over{2\Delta\sqrt{x}}}dx\, .
\ee
In these coordinates the 5D MP metric is
\[
ds^2=-dv^2
+(r^2+a^2)\sin^2\theta d\tilde{\phi}^2+(r^2+b^2)\cos^2\theta
d\tilde{\psi}^2
\]
\be\n{6.4}
+\rho^2d\theta^2+{1\over{\rho^2}}\left(dv+a\sin^2\theta d\tilde{\phi}+b\cos^2\theta
d\tilde{\psi}\right)^2
\ee
\[
+ 2dr\left(dv+a\sin^2\theta d\tilde{\phi}+b\cos^2\theta
d\tilde{\psi}\right)\, .
\]
It is  more convenient to use again the radial coordinate $r$ instead
of $x=r^2$. (We still use units in which $r_0=1$.) The original Killing
vectors $\partial_t$, $\partial_{\phi}$ and  $\partial_{\psi}$ take
the form $\partial_v$, $\partial_{\tilde{\phi}}$ and 
$\partial_{\tilde{\psi}}$.

The principal Killing string in these coordinates has a  simple
from
\be\n{6.5}
\theta=\theta_0\, ,\hspace{0.3cm}
\tilde{\phi}=\tilde{\phi}_0\,  ,\hspace{0.3cm}
\tilde{\psi}=\tilde{\psi}_0\, .
\ee 
We use  coordinates of $\zeta^0=v$ and $\zeta^1=r$ as
coordinates on $\Sigma$. The induced metric in these coordinates is
\be\n{6.6}
d\gamma^2=-F\,dv^2+\,2dr dv\,,
\ee
\be\n{6.7}
F=1-{1\over r^2+P_0^2}\, ,\hspace{0.3cm}P_0^2=a^2\cos^2\theta_0+b^2\sin^2\theta_0\,.
\ee
This is a metric of a 2D black hole with an event horizon located at 
\be\n{6.8}
r^2+P^2_0=1\, .
\ee
The surface gravity  is given by
$\kappa={1\over{2}} F_{,r}$, 
evaluated at the horizon $F=0$. We get 
\be\n{6.9}
\kappa=\sqrt{1-P_0^2}.
\ee
For comparison, we restate the five dimensional surface gravity
$\kappa_{(5)}$ from (\ref{3.5}) in an explicit form
\be\n{6.10}
\kappa_{(5)}=\sqrt{2}\, {\sqrt{B^2-4a^2b^2}\over{\sqrt{B+\sqrt{B^2-4a^2b^2}}}}\,
,
\ee
where $B=1-a^2-b^2$. Denote
\be\n{6.11}
f(r)={2r^2-B\over r}\, .
\ee
This function is monotonically increasing. One also has
\be\n{6.12}
\kappa_{(5)}=f(r_+)\, ,
\ee
where $r_+^2={1\over 2}(B+\sqrt{B^2-4a^2b^2})$.
Since $r_+\le B^{1/2}$ one has
\be\n{6.13}
\kappa_{(5)}=f(r_+)\le f(B^{1/2})=B^{1/2}\le \kappa_{(2)}\, .
\ee
This relation shows that in a general case the surface gravity of 2D
principal Killing string hole is greater than the surface gravity of
the 5D MP black hole. The equality is possible only if
\be\n{6.14}
a^2\sin^2\theta_0+b^2\cos^2\theta_0=0\, .
\ee
This occurs either when $a=b=0$ and the black hole is non-rotating, or
when one of the rotation parameters, say $b$,  vanishes and the string
is in the $\theta=0$ plane.

\section{Higher dimensional rotating black holes attached
to a string}

\setcounter{equation}0

\subsection{Higher dimensional MP metric}

We demonstrate now that the  results concerning principal
Killing strings can be generalized to the higher dimensional case, that
is when the number $N$ of spatial dimensions is greater than 4.
Corresponding MP metrics  \cite{MyPe:86} have slightly different form
for $N$ even and odd.

For an even number of spatial dimensions
\[
ds^2=-\,{dt^2}+\sum_i {(r^2+a_i^2)}{(d\mu_i^2+{\mu_i^2}d\phi_i^2)}+
{{\Pi L}\over{\Pi-m  r^2}}dr^2
\]
\be\n{7.1}
+{{m  r^2}\over{\Pi L}}(dt+\sum_i a_i
\mu_i^2 d\phi_i)^2\, ,
\ee
where  $\Pi$ and $L$ given by
\be\n{7.2}
\Pi=\prod_{i}(r^2+a_i^2)\,,\hspace{0.3cm}
L=1-\sum_i{{a_i^2\mu_i^2}\over{r^2+a_i^2}}\, ,
\ee
and $\mu_i$ obey the relation
\be\n{7.3}
\sum_i \mu_i^2=1\, .
\ee

For an odd number of spatial dimensions we instead have
\[
ds^2=-\,{dt^2}+r^2 d\mu ^2+\sum_i
(r^2+a_i^2)(d\mu_i^2+{\mu_i^2}d\phi_i^2) 
\]
\be\n{7.4}
+{\Pi L\over{\Pi-m  r}}dr^2 +{m  r\over{\Pi L}}
(dt+\sum_i{a_i\mu_i^2 d\phi_i})^2 \, .
\ee
\be\n{7.5}
\mu^2+\sum_i \mu_i^2=1\, .
\ee
Here and later it is assumed that the summation $\sum_i$ is taken
from $i=1$ to $i=[N/2]$, where $[A]$ means the integer part of $A$.

The parameter $m=r_0^{N-2}$, $r_0$ being the gravitational radius, is
related to the mass $M$ of the black hole as follows
\be\n{7.6}
M={(N-1){\cal A}_{N-1}\over 16\pi G_{N+1}} r_0^{N-2} \, ,
\ee
where
\be\n{7.7}
{\cal A}_{N-1}={2\pi^{N/2}\over \Gamma(N/2)}
\ee 
is the area of a unit sphere $S^{N-1}$ and $G_{N+1}$  is the
$N+1$-dimensional gravitational coupling constant which has
dimensionality $[\mbox{length}]^{(N-2)}/[\mbox{mass}]$. The angular
momenta $J_i$ of the black hole are defined as
\be\n{7.8}
J_i=-{{\cal A}_{N-1}\over 8\pi G_{N+1}} ma_i=-{2\over N-1} Ma_i\, .
\ee
(Note that the sign of the rotation parameters in the MP metric is
opposite to the one adopted in the Kerr metric.)

The principal null vectors $l_{\pm}$  are given by
\be\n{7.9}
l_{\pm}^{\mu}\pr_{\mu}=\Lambda \left(\pr_t 
-\sum_i {a_i \over{r^2+a_i^2}}\pr_{\phi_i}\right)\pm \pr_r\, .
\ee
where 
\ba
\Lambda&=&\Pi/(\Pi-m  r)\, ,\, \, \mbox{for odd} \, \, N\, ; \\
\Lambda&=&\Pi/(\Pi-m  r^2)\, ,\, \, \mbox{for even}\, \, N\, .
\ea

Let us introduce the ingoing ($-$) and outgoing ($+$)
Eddington-Finkelstein coordinates $(v_{\pm},r,\tilde{\phi}_{\pm i})$
\be\n{7.11}
dv_{\pm}=dt\mp \Lambda dr\, ,
\ee
\be\n{7.12}
d\tilde{\phi}_{\pm i}=d\phi_i\pm{\Lambda a_i \over{r^2+a_i^2}}dr\, .
\ee
One has
\be\n{7.13}
l_{\pm}^{\mu}\pr_{\mu}=\pm\pr_{r}\, ,
\hspace{0.5cm}l_{\pm \mu}dx^{\mu}=-[dv_{\pm}+\sum_i \mu_i^2 a_i
d\tilde{\phi}_{\pm i}]\, .
\ee
For odd $N$  the MP metric takes the form
\[
ds^2=-dv_{\pm}^2+\sum_i (r^2+a_i^2)(d\mu_i^2+\mu_i^2 d\tilde{\phi}_{\pm i}^2)
\]
\be\n{7.14}
+{m r \over{\Pi L}}(l_{\pm \mu}dx^{\mu})^2 \pm 2dr(l_{\pm
\mu}dx^{\mu})+r^2 d\mu^2 \, .
\ee
For even $N$ the metric is similar. The only difference is that 
the last term is absent and $m r$ must be changed to $m r^2$.

In the MP metrics (\ref{7.1}), (\ref{7.4}) and (\ref{7.14}) the
quantities $\mu$ and $\mu_i$ are not independent. They obey the
restrictions (\ref{7.3}) or (\ref{7.5}) indicating that these
quantities belong to a unit sphere. We denote by $\theta_a$
($a=1,\ldots,[(N-1)/2]$) independent coordinates on the sphere. We
shall also use a notation $\omega_{m}=(\theta_a,\tilde{\phi}_i)$
($m=(2,\ldots,N)$) for a total set of `angular' coordinates. 

Denote as earlier by $\xi$ a timelike at infinity
Killing vector, $\xi^{\mu}\partial_{\mu}=\partial_t$. Then the
principal null vectors are eigen-vectors of $\xi_{\mu ;\nu}$, 
\be\n{7.16}
\xi_{\mu ;\nu}l_{\pm}^{\nu}=\pm {1\over{2}}\pr_r F\, l_{\pm\mu}\, ,
\ee
where  $F=-\xi^2=-g_{vv}$.
This is a generalization of the property (\ref{5.8}) to the higher
dimensional case. To prove this relation we consider  an expression
\be\n{7.17}
\xi_{\mu ;\nu}l_{\pm}^{\nu}=\xi_{\mu ,\nu}l_{\pm}^{\nu}
-\Gamma_{\sigma,\mu\nu}\xi^{\sigma}l_{\pm}^{\nu}\, .
\ee
Since  $l_{\pm}^{\nu}=\pm\delta^{\nu}_r$ and
$\xi^{\mu}=\delta^{\mu}_v$, we obtain
\[
\xi_{\mu ;\nu}l_{\pm}^{\nu}=\pm g_{\mu v ,r}\mp {1\over{2}}(g_{v \mu ,r}
+g_{vr,\mu}-g_{r\mu ,v})\, ,
\]
\be\n{7.18}
=\pm {1\over{2}}g_{v \mu ,r}=\pm {1\over{2}}F_{,r} l_{\pm\mu}\, .
\ee
Combining relations (\ref{7.18}) and (\ref{7.19}) one proves
the relation (\ref{7.16}).

\subsection{Higher-dimensional principal Killing strings}

Now we show that a 2D surface $\Sigma$ spanned by the vectors
$l_{\pm}^{\mu}$ and $\xi^{\mu}$ is a solution to the Nambu-Goto equations
of motions, and is thus a minimal surface. We use coordinates
$\zeta^0=v$ and $\zeta^1=r$ to parametrize $\Sigma$. The worldsheet
equation is
\be\n{7.19}
X^0=v\, ,\hspace{0.2cm}
X^1=r\, ,\hspace{0.2cm}
X^m=\omega_m=\mbox{const}\, .
\ee
We write the Nambu-Goto equations as follows
\be\n{7.20}
g_{\mu\nu}\Box X^{\nu}+
G^{AB}\Gamma_{\mu,\alpha\beta}X^{\alpha}_{,A}X^{\beta}_{,B}=0\, ,
\ee
with
\be\n{7.21}
\Box X^{\nu}={1\over{\sqrt{-G}}}\pr_{A}(\sqrt{-G}G^{AB}\pr_{B}X^{\nu})\, ,
\ee
\be\n{7.22}
G^{AB}\pr_{A}\pr_{B}=F\pr_{r}^2\mp 2\pr_r \pr_v\, ,\hspace{0.3cm}
\sqrt{-G}=1\, .
\ee
The first term in (\ref{7.20}) is
\be\n{7.23}
g_{\mu\nu}\Box X^{\nu}=g_{\mu\nu}[\pr_r (F \pr_r X^{\nu}) \mp 2\pr_r \pr_v
X^{\nu}]=g_{\mu r} \pr_r F\, ,
\ee
and the second term reads
\[
G^{AB}\Gamma_{\mu,\alpha\beta}X^{\alpha}_{,A}X^{\beta}_{,B}=
\Gamma_{\mu,\alpha\beta}[F X^{\alpha}_{,r}X^{\beta}_{,r}
\mp 2X^{\alpha}_{,r}X^{\beta}_{,v}]
\]
\be\n{7.24}
={1\over{2}}F (2g_{\mu r ,r}-g_{rr,\mu})\mp (g_{\mu r ,v}
+ g_{\mu v ,r}- g_{vr,\mu})\, .
\ee
The form of the metric (\ref{7.14}) implies that 
\be\n{7.25}
g_{\mu r ,r}=g_{r r}=g_{\mu r ,v}=g_{v r ,\mu}=0\, .
\ee
Thus the left hand side of (\ref{7.20}) is
\be\n{7.26}
g_{\mu r} \pr_r F \mp g_{v \mu ,r}= \pm l_{\pm\mu}\pr_{r} F \mp F_{,r}l_{\pm\mu}=0\, .
\ee
This result means that the principal Killing surface (\ref{7.19}) is a
solution of the Nambu-Goto equations (\ref{7.20}). A principal Killing
string which is regular at the future event horizon is generated by
$l_-$.

\section{Interaction of a higher dimensional rotating black hole with
a string}
\setcounter{equation}0

Until now we considered a cosmic string as a test object and neglect its
action on the black hole. We show now that the interaction between the
string and the black hole results in the transfer of the angular
momenta from the black hole to the string. 

Suppose there exists a distribution of matter outside the rotating
black hole. Then the fluxes of  energy and
angular momenta of this matter through a surface $r=$const are
\be\n{8.1}
\Delta E=-\int T_{\mu}^{\, \nu}\, \xi^\mu d\sigma_\nu \,, 
\hspace{0.2cm}
\Delta J_i=\int T_{\mu}^{\, \nu}\, \xi_{i}^\mu d\sigma_\nu \,. 
\ee
Here $T_{\mu\nu}$ is the stress-energy tensor of the matter, 
and $\xi_i^{\mu}\pr_{\mu}=\pr_{\tilde{\phi}_i}$ are the Killing
vectors connected with the rotational invariance of the MP metric. 
The expression for $d\sigma_{\mu}$ can be written as follows
\be\n{8.2}
d\sigma_{\mu}=r_{,\mu}\sqrt{-g}\, dv\, d\omega^{N-1}\, ,
\ee 
where
\be\n{8.3}
d\omega^{N-1}=\prod_{m=1}^{N-1} d\omega_m\, .
\ee

For a stationary configuration the (constant) rate of energy  and 
angular momentum fluxes from the black hole through the $r=$const
surface are
\be\n{8.4}
\dot{E}\equiv {dE\over dv}=- \int  d\omega^{N-1}\, \sqrt{-g}\, 
T_{\mu}^{\, \nu}\, \xi^\mu d\sigma_\nu\, ,
\ee
\be\n{8.5}
\dot{J}_i\equiv {dJ_i\over dv}= \int  d\omega^{N-1}\, \sqrt{-g}\,
T_{\mu}^{\, \nu}\, \xi_{i}^\mu d\sigma_\nu\, .
\ee
If a part of initially infinite string is captured by a black hole its
world-sheet in the black hole exterior consists of two segments. For a
stationary cosmic string each of these segments is a principal Killing
string. We calculate the energy and angular momenta transfer for one
segment. The stress-energy  tensor of the string is
defined as follows:
\be\n{8.6}
\sqrt{-g}T^{\mu\nu}=-\mu^* \int d^{2}\zeta \delta^{(N+1)}(X-X(\zeta))\,
t^{\mu\nu}\, ,
\ee
\be\n{8.7}
t^{\mu\nu}=\sqrt{-G} G^{AB}X^{\mu}_{,A}X^{\nu}_{,B}~~\, , 
\ee
where $\mu^*$ is the tension of the string. For the principal Killing
string one has
\be\n{8.8}
t^{\mu\nu}=F\, l^{\mu}\, l^{\nu}-2l^{(\mu}\xi^{\nu)}\, .
\ee
Since we are considering strings which are regular at the future event
horizon, $l$ is equat to $l_-$. Using (\ref{7.14}) one obtaines
\be\n{8.9}
t_{\mu}^{\nu}\xi^{\mu}=\xi^{\nu}\, ,
\hspace{0.2cm}
t_{\mu}^{\nu}\xi_{i}^{\mu}=a_i\, \mu_i^2\, (\xi^{\nu}-l^{\nu})\, .
\ee

Substituting (\ref{8.6}) into (\ref{8.4}) and (\ref{8.5}), taking
the integrals and using the relations (\ref{8.9}) one finds
\be\n{8.10}
\dot{E}=0\, ,
\ee
\be\n{8.11}
\dot{J_i}=-\mu^* a_i \mu^2_i\, .
\ee
The angular momentum flux does not depend on $r$. This is in
accordance with the conservation law. The energy flux vanishes.
The angular momenta transfer to the string results in the decrease of
the corresponding angular momenta of the black hole
\be\n{8.12}
\dot{J}^{BH}_i=-\dot{J_i}\,.
\ee
Using the expression (\ref{7.8}) for the angular momenta of the black
hole one can obtain the following equation for the loss of the
angular momenta of the black hole
\be\n{8.13}
\dot{J}^{BH}_i=- \mu^2_i (N-1){\mu^* \over{M}} J^{BH}_i\, .
\ee
In this equation we took into account that there are 2 string
segments attached to the black hole. We choose the second segment to
be an inverse image   of the first one with respect to the center of
the black hole so that for it $\tilde{\phi}\to \pi+\tilde{\phi}$ (and
$\mu\to -\mu$ for  odd $N$).
This guarantees that under the action of the string
the black hole remains at rest as a whole. In 4D case of the Kerr
black hole the obtained result coincides with result of \cite{FFS}.

The equation (\ref{8.13}) shows that $\dot{J}^{BH}_i=0$ either when
$J^{BH}_i=0$, that is when the black hole does not rotate in the
$i$-th bi-plane of rotation, or when $\mu_i^2=0$. In the 4D case of
the Kerr black hole the latter condition implies that the string is
directed along the axis of rotation of the black hole.

Equation (\ref{8.13}) shows that the bulk components of the angular
momentum $J_i$ with $\mu_i\ne 0$ decrease. In the case when
$\mu^*$ is small ($\mu^* r_+/M\ll 1$) this process is very slow and
one can use a quasistationary approximation, that is to consider it as a
slow change from one stationary configuration to another one. In this
approximation the evolution of the system can be described as a
evolution in the space of parameters characterizing quasistationary
system black-hole--cosmic-string. 

To estimate the characteristic time of the slowing down of the bulk
components of the black hole rotation let us consider a case when
only one component of the initial angular momentum is non-vanishing
and the corresponding $\mu_i=1$. In this case all other $\mu_i$ vanish.
This is a higher dimensional analogue of a cosmic string in the
equatorial plane of the 4D Kerr black hole. Since $\mu^*$ is small
and the evolution is adiabatic, the surface area of the black hole 
\be\n{8.14}
{\cal A}=r_+^{N-3}(r_+^2+a^2)A_{N-1}\, ,
\ee
remains constant. Here $r_+$ is a position of the event horizon
defined by the relation
\be\n{8.15}
{16\pi G_{N+1}\over N-1} M =r_+^{N-4}(r_+^2+a^2)A_{N-1}\, .
\ee
By using the equations (\ref{8.13}), (\ref{8.14}) and (\ref{7.8}) one
gets
\be\n{8.16}
\dot{r}_+=\frac{32 \pi G_{N+1}
\mu^*}{(N-1)A_{N-1}}\frac{a^2}{r_+^{N-5}(r^2_+ +a^2)^2}\, .
\ee
This equation shows that during the evolution $r_+$ does not decrease
and it remains constant only when $a=0$. Let us rewrite the relation
(\ref{8.14}) in the form
\be\n{8.17}
\beta\equiv {a^2\over r_+^2}=\frac{\mathcal{A}}{A_{N-1}r_+^{N-1}}-1\,
.
\ee 
This relation shows that $r_+$ grows until it reaches its final value
\be\n{8.18}
r_f=\left( \frac{\mathcal{A}}{A_{N-1}}\right)^{1/(N-1)}\, .
\ee
Using (\ref{8.18}) and (\ref{8.16}) one can obtain the following
equation which defines the evolution of the rotation parameter
$\beta$
\be\n{8.19}
\dot{\beta}=-{1\over T}\frac{\beta}{(1+\beta)^{1\over N-1}}\, ,
\ee
where
\be\n{8.20}
T={M_f \over   2(N-1)\mu^*}\, ,
\ee
where $M_f={\cal A} (N-1)/(16\pi G_{N+1}r_f)$ is the final value of
mass of the black hole. The parameter $T$ has dimensionality of time
and it determines the characteristic time scale of the process of
slowing down the rotation of the black hole.

The equation (\ref{8.19}) can be solved analytically. The solution is
\be\n{8.21}
\ln{\beta}+\frac{\beta\,\, {}_3 F_2(1,1,{N-2\over
N-1};2,2;-\beta)}{N-1}=-\frac{t}{T}+C\, , 
\ee
where $C$ is the constant of integration. The hypergeometric function
takes the value 1 when $\beta=0$. In the limit $t\to\infty$ we have
$\beta\to 0$ so  the logarithm is the leading term in the left hand
side. Thus at late time the function $\beta$ has the following
asymptotic form
\be\n{8.22}
\beta=\beta_0\exp\left(-t/T\right)\, .
\ee

\section{Discussions}
\setcounter{equation}0

Let us summarize the results obtained in the paper. We considered
interaction of a cosmic string with rotating higher dimensional black
holes. In 5D case the stationary string equations allow separation of
variables. This occurs because of the existence of a sufficient
number of integral of motions which, in particular, include the
conservation law connected with the Killing tensor. Increasing the
number of spacetime dimensions from 5 to 6 does not produce a new
Killing vector. Moreover it is probable that the 6D MP spacetime does
not have the Killing tensor. Lack of the sufficient number of
integrals of motion in six and higher number of spacetime
dimensions  make it improbable the separation of variables of
stationary string equations in such higher dimensional spacetimes.

Among all stationary solutions of string equations there is a special
class, principal Killing strings. Their characteristic property is
that such solutions describe stationary strings which starting from
spatial infinity cross the ergosphere and enter the horizon and
remain regular. We demonstrated that these solutions exist in an
arbitrary number of dimensions. Internal geometry of these principal
Killing strings is a geometry of 2D static black hole with its 2D
horizon located at the intersection of the string with infinite
red-shift surface. Perturbations propagating along the string cannot
escape the region inside the 2D horizon. At the same time, the causal
signals propagating in the bulk space can transfer the information
about the state of the string inside its 2D horizon up to the point
of its intersection with the horizon of the higher-dimensional black
hole. A 2D observer would interpret this as extracting the
information from the 2D black hole by means of extra dimensions.

An interaction of a rotating black hole with the string results in the
reduction of some components of its angular momenta.
In a general case there exists an angular
momentum flux from a black hole to the attached string.
The flux of the angular momentum is proportional to the
tension of the string $\mu^*$. The
characteristic time of the relaxation process during which the black
hole reaches its final state is given by relation (\ref{8.20}).
In the first order in the string tension $\mu^*$ there is no energy
flux. The final stationary black hole may have rotation only in those
planes for which $\mu_i=0$.

%\newpage
\bigskip

\vspace{12pt} {\bf Acknowledgments}:\ \  This work was partly
supported  by  the Natural Sciences and Engineering Research Council
of Canada. One of the authors (V. F.) is grateful to the Killam Trust
for its financial support.

\bigskip

\appendix

\section{Uniqueness property}
\setcounter{equation}0

We demonstrate now that a principal Killing surface is the only 
possible worldsheet of a stationary string which crosses the infinite
red-shift surface and remain regular. 

Consider a stationary surface $\Sigma$. It can be presented as a one
parameter family of trajectories of the Killing vector $\xi$. Let $v$
be a Killing time parameter, then $d\zeta^{A}/dv=\xi^{A}$.   We use
$v$ as one of the coordinates on $\Sigma$. We use an ambiguity in the
choice of the other coordinate, $r$, to put $G_{rr}=0$. For this
choice
\be\n{a.1}
d\gamma^2=-F\,dv^2+\,2\alpha dr dv\,,
\ee
\be\n{a.2}
F=-\xi^2\, ,\hspace{0.3cm}
\alpha=-(\xi\cdot l)
\, ,\hspace{0.3cm}
l^{A}\partial_A=-\partial_r\, .
\ee
The Killing vector $\xi$ to gether with a null vector $l$ span
$\Sigma$. It is easy to check that $\xi$ is also a Killing vector for
the induced metric $G_{AB}$. Using Killing equation $\xi_{(A:B)}=0$ in the
induced metric one has $\partial_v F=\partial_v \alpha=0$. Using an
ambiguity $r\to f(r)$ one can always put $\alpha=1$. Denote by $n_R$ a
complete set of mutually orthogonal unit vectors orthogonal to
$\Sigma$. Then a definition of the second fundamental form
(\ref{5.16}) implies that relations (\ref{5.17})-(\ref{5.19}) are
still valid
\be\n{a.3}
\Omega_{R}=(n_R\cdot z)\, ,
\ee 
\be\n{a.4}
z^{\mu}=-2\xi^{\mu}_{\ \ ;\rho}l^{\rho}+Fl^{\rho}l^{\mu}_{\,\,;\rho}\,
.
\ee
Since $\xi_{\mu;\nu}$ is antisymmetric and $l$ is null one also has
\be\n{a.5}
(l\cdot z)=0\, .
\ee

The surface $\Sigma$ is minimal if $\Omega_R=0$. For such a surface
\be\n{a.6}
\Omega^2=\Omega_R\Omega^R=(g^{\mu\nu}-G^{AB} x_{,A}^{\mu}
x_{,B}^{\nu}) z_{\mu} z_{\nu}=0\, .
\ee
Using (\ref{a.5}) it is easy to check that for our choice of the
coordinates
\be\n{a.7}
G^{AB} x_{,A}^{\mu} x_{,B}^{\nu} z_{\mu} z_{\nu}=(l\cdot z)[2(\xi\cdot
z)+F(l\cdot z)]=0\, .
\ee
Thus $z$ is a null vector, $g_{\mu\nu}z^{\mu}z^{\nu}=0$. Equation
(\ref{a.5}) implies that 
\be\n{a.8}
z^{\mu}=q l^{\mu}\, .
\ee
To determine $q$ we multiply this relation by $\xi_{\mu}$
\be\n{a.9}
q=l^{\rho}(\xi^2)_{,\rho}+F\xi_{\nu;\rho}l^{\nu}l^{\rho}={dF\over
dr}\, .
\ee

Using (\ref{a.4}) and (\ref{a.8}) we have
\be\n{a.10}
2\xi^{\mu}_{\ \ ;\rho}l^{\rho}=Fl^{\rho}l^{\mu}_{\,\,;\rho}-{dF\over
dr}l^{\mu}\, .
\ee
This relation shows that at the infinite red-shift surface, $F=0$,
the null vector $l$ is the eigenvector of $\xi^{\mu}_{\ \ ;\rho}$
with the eigenvalue $-F_{,r}/2$ and hence it coincides with $l_-$. In
the vicinity of the infinite red-shift surface one has
\be\n{a.11}
l=(1+\lambda) l_- +\mu\bar{m}+\bar{\mu} m+\nu k\, ,
\ee
The term proportional to $l_+$ does not appear since $l\cdot l=0$.
Equations (\ref{5.3})--(\ref{5.5}) implies that
\be\n{a.12}
(m\cdot \xi)={i\sin\theta\, \cos\theta \over \sqrt{2}\rho\,
P}(a^2-b^2)\, ,
\ee
\be\n{a.13}
(k\cdot \xi)=-{ab\over \sqrt{x}P}\, .
\ee
Using these relations and $\xi\cdot l=-1$ we obtain
\be\n{a.14}
\lambda= {1\over{P  }}
\left[{i(a^2-b^2)\sin\theta\cos\theta\over{\rho\sqrt{2}}}(\bar{\mu}-\mu)
-{ab\over{\sqrt{x}}}\nu\right]\, ,
\ee
where $P$ is defined by (\ref{3.2}). Vectors $m$ and $k$ are
given by (\ref{5.3}) and (\ref{5.5}), respectively.  In our
perturbation analysis we keep only those terms which are of the 
first order in $\mu$ and $\nu$, and drop anything that is a multiple
of these two.

If we contract (\ref{a.10}) with $\bar{m}_{\mu}$ and use the
relations
\be\n{a.15}
\bar{m}_{\mu}l^{\rho}\xi^{\mu}_{\,\,\,\, ;\rho}
=-{i\bar{\mu}(1-F)P   \over{\rho^2}}\, ,
\ee
\be\n{a.16}
\bar{m}_{\mu} l^{\mu}= \bar{\mu}\, ,
\ee
\be\n{a.17}
\bar{m}_{\mu} l^{\rho}l^{\mu}_{\,\,\,\, ;\rho}
= l_-^{\rho}\bar{\mu}_{,\rho}+
{\bar{\mu}\over{\rho^2}}\left(2iP  -\sqrt{x}\right)\, ,
\ee
we arrive at
\be\n{a.18}
Fl_-^{\rho}\bar{\mu}_{,\rho}=-F{d\bar{\mu}\over{dr}}=-\Omega\bar{\mu},
\ee
where
\be\n{a.19}
\Omega={2iP  -\sqrt{x} F\over{\rho^2}}-{dF\over{dr}}\, .
\ee
So if we define the tortoise coordinate $r^*$ as
\be\n{a.20}
{dr\over{dr^*}}=F\, ,
\ee
so that the ergosphere lies at $r*\rightarrow -\infty$, then we can
solve for $\bar{\mu}$ as
\be\n{a.21}
\bar{\mu}=\bar{\mu}_0 e^{\Omega r^*}\, ,
\ee
and since $Re(\Omega)<0$ we must have $\bar{\mu}_0 =0$ to ensure that
the solution is regular at the ergosphere.

If we contract (\ref{a.10}) with $k_{\mu}$ and use  the following
relations
\be\n{a.22}
k _{\mu}l^{\rho}\xi^{\mu}_{\,\,\, ;\rho}=0\, ,
\ee
\be\n{a.23}
k _{\mu}l^{\mu}=\nu\, ,
\ee
\be\n{a.24}
k _{\mu}l^{\rho}l^{\mu}_{\,\,\, ;\rho}=-{\nu\over{\sqrt{x}}}
+l_-^{\rho}\nu_{,\rho}\, ,
\ee
we arrive at
\be\n{a.25}
Fl_-^{\rho}\nu_{,\rho}=-F{d\nu\over{dr}}=-W\nu \,,
\ee
where
\be\n{a.26}
W=-{dF\over{dr}}-{F\over{\sqrt{x}}}\, .
\ee
so we can get
\be\n{a.27}
\nu=\nu_0 e^{W r^*}\, .
\ee
Again since $Re(W)<0$ we must have $\nu_0 =0$ to have a solution
that is regular at the ergosphere. 

We demonstrated that there is no a regular stationary string solution
which coincides with $l_-$ at the infinite red-shift surface but
differs from it outside the ergosphere. This completes the proof.

%%%%%%%%%%%%%%%%%%%%%%%%%%%%%%%%%%%%%%%%%%%%%%%%%%%%%%%%%%%%%%%%%%%%%%%%%%%%%

%%%%%%%%%%%%%%%%%%%%%%%%%%%%%%%%%%%%%%%%%%%%%%%%%%%%%%%%%%%%%%%%%%%%%%%%%%%


\begin{thebibliography}{9}

\bibitem{Chandra} S. Chandrasekhar, {\em The Mathematical Theory of
Black Holes}, (Oxford: Clarendon) (1983). 

\bibitem{Cart:67} B. Carter,  Phys. Rev. {\bf 174}, 1559 (1967).

\bibitem{Cart:68} B. Carter,  Commun. Math. Phys. {\bf 10}, 280 (1968).

\bibitem{Teuk:68} S. Teukolsky,  Phys. Rev. Lett. {\bf 29}, 1114 (1968).

\bibitem{FrSkZeHe:89}  V. P. Frolov ,  V. D. Skarzhinsky,  A. I. Zel'nikov 
and O. Heinrich,  Phys. Lett. {\bf B224}, 255 (1989). 

\bibitem{CaFr:89} B. Carter and V. P. Frolov,  Class. Quantum Grav.
{\bf 6}, 569 (1989).

\bibitem{MyPe:86} R. C. Myers and M.~J. Perry, Ann. Phys. {\bf 172},  304
(1986). 

\bibitem{FrSt:03a} V. Frolov and  D. Stojkovic, Phys.Rev. {\bf D68},
064011 (2003). 

\bibitem{FrSt:03b} V. Frolov and D. Stojkovic, Phys.Rev. {\bf D67},
084004 (2003). 

\bibitem{FrHeLa:96} V. Frolov, S. Hendy and  A. L. Larsen, Phys.Rev.
{\bf D54}, 5093 (1996).

\bibitem{Geroch} R. Geroch, J. Math. Phys. {\bf 12}, 918 (1971).


\bibitem{CGL} G. Clement, D. Gal'tsov and C. Leygnac, Phys.Rev. {\bf
D67}, 024012 (2003). 


\bibitem{MTW} C.~W.~Misner, K.~S.~Thorne, and J.~A.~Wheeler, {\em
Gravitation} (W.H.~Freeman, San Francisco, 1973).


\bibitem{FFS} V. Frolov, D. Fursaev, and D. Stojkovic; 
e-Prints:gr-qc/0403002 and gr-qc/0403054  (2004).

\end{thebibliography}
\end{document}